\documentclass[aps,prb,twocolumn,superscriptaddress,nobibnotes,amsmath,amssymb,reprint]{revtex4-1}

\usepackage{graphicx}
\usepackage{color}
\usepackage[normalem]{ulem}

\newcommand{\mysub}[2]{#1_{\text{#2}}}
\newcommand{\myvareq}[4]{$#1_{\text{#2}}=#3\,\mathrm{#4}$}
\newcommand{\myq}[2]{$#1\,\mathrm{#2}$}

\begin{document}
\title{Spatial and energy resolution of electronic states by shot noise}
\author{E.S.~Tikhonov}
\email[e-mail:]{tikhonov@issp.ac.ru}
\affiliation{Institute of Solid State Physics, Russian Academy of Sciences, 142432 Chernogolovka, Russian Federation}
\author{A.O.~Denisov}
\altaffiliation{present address: Department of Physics, Princeton University, Princeton, New Jersey 08544, USA}
\affiliation{Institute of Solid State Physics, Russian Academy of Sciences, 142432 Chernogolovka, Russian Federation}
\author{S.U.~Piatrusha}
\affiliation{Institute of Solid State Physics, Russian Academy of Sciences, 142432 Chernogolovka, Russian Federation}
\author{I.N.~Khrapach}
\affiliation{Institute of Solid State Physics, Russian Academy of Sciences, 142432 Chernogolovka, Russian Federation}
\affiliation{Moscow Institute of Physics and Technology, 141700 Dolgoprudny, Russia}
\affiliation{Russian Quantum Center, 121205 Skolkovo, Moscow, Russia}
\author{J.P.~Pekola}
\affiliation{QTF Centre of Excellence, Department of Applied Physics, Aalto University, FI-00076 Aalto, Finland}
\affiliation{Moscow Institute of Physics and Technology, 141700 Dolgoprudny, Russia}
\author{B.~Karimi}
\affiliation{QTF Centre of Excellence, Department of Applied Physics, Aalto University, FI-00076 Aalto, Finland}
\author{R.N.~Jabdaraghi}	
\affiliation{QTF Centre of Excellence, Department of Applied Physics, Aalto University, FI-00076 Aalto, Finland}
\affiliation{VTT Technical Research Centre of Finland Ltd}
\author{V.S.~Khrapai}
\affiliation{Institute of Solid State Physics, Russian Academy of Sciences, 142432 Chernogolovka, Russian Federation}
\affiliation{National Research University Higher School of Economics, 20 Myasnitskaya Street, Moscow 101000, Russia}
\begin{abstract}
Shot noise measurements are widely used for the characterization of nonequilibrium configurations in electronic conductors. The recently introduced quantum tomography approach was implemented for the studies of electronic wavefunctions of few-electron excitations created by periodic voltage pulses in phase-coherent ballistic conductors based on the high-quality GaAs two-dimensional electron gas. Still relying on the manifestation of Fermi correlations in noise, we focus on the simpler and more general approach beneficial for the local measurements of energy distribution (ED) in electronic systems with arbitrary excitations with well-defined energies and random phases. Using biased diffusive metallic wire as a testbed, we demonstrate the power of this approach and extract the well-known double-step ED from the shot noise of a weakly coupled tunnel junction. Our experiment paves the way for the local measurements of generic nonequilibrium configurations applicable to virtually any conductor.
\end{abstract}
\maketitle
Nanoscale temperature mapping and control of nonequilibrium configurations has attracted much interest recently. The prominent examples range from thermometry in a living cell~\cite{Kucsko2013} to thermal imaging of quantum systems~\cite{Halbertal2016} and nanoscale devices~\cite{Menges2016}. Along with direct thermal measurements and NVC- and SQUID-based thermometers~\cite{Kolkowitz2015}, primary shot noise thermometry is also attractive due to its self-calibrating nature~\cite{Spietz2003,Kemiktarak2007}. Historically, it was first used for the study of hot-electron regime in metallic resistors~\cite{PhysRevLett.55.422,PhysRevLett.76.3806,PhysRevB.59.2871} and was later on extended to primary electronic thermometry~\cite{Spietz2003} and to the studies of graphene~\cite{PhysRevLett.109.056805,Betz2012,PhysRevX.3.041008,PhysRevB.91.121414,PhysRevB.93.075410,PhysRevB.99.075419}. 

Shot noise power of the current fluctuations~$\mysub{S}{I}$ in a dc-biased two-terminal conductor, however, doesn't provide neither local nor energy resolution. The reason is that random fluctuations of the occupation numbers of the electronic quantum states are averaged both in the energy interval where electron scattering is possible, and along the length of the device~\cite{Nagaev1992}. This fundamental constraint set by current conservation~\cite{BLANTER20001}, makes accessible only the device-averaged nonequilibrium noise temperature~$\mysub{T}{N}$. To gain further insight into the charge kinetics, one can measure the dynamical response of noise to an ac excitation~\cite{PhysRevLett.72.538,PhysRevLett.80.2437,PhysRevB.87.075403,PhysRevLett.116.236601}, or implement the special design of the experiment to guide currents in a magnetic field~\cite{PhysRevLett.112.166801}.

Recently, shot noise was utilized to study electronic wavefunctions emitted by the time-dependent currents in a phase-coherent conductor~\cite{Jullien2014,Bisognin2019}. Essentially, the implemented tomographic approach uses antibunching of fermions due to Pauli principle in the geometry of a beam splitter. So far, it was used for characterization of excitations created by specific $T$-periodic identical voltage pulses. These excitations are a coherent superposition of single-particle eigenstates~\cite{PhysRevLett.97.116403} with different energies in the range of $\sim\hbar/T$ with a spatial extension of $v_FT$. On the other hand, the fermionic system with arbitrary excitations with well-defined energies and random phases should rather be itself characterized by the energy distribution (ED). As we experimentally demonstrate below, alternative approach~\cite{PhysRevB.60.2375}, still relying on the manifestation of Fermi correlations in noise, is beneficial for its local measurements.

Up to now, ED measurements in mesoscale devices typically relied on the spectral sensitivity of the used detector. 
This spectral sensitivity inherent, e.g., to superconducting electrodes or quantum dots (QD) with discrete electronic levels, allowed to use them as sensors for the measurements of ED inside current-driven mesoscopic metallic wires~\cite{PhysRevLett.79.3490,PhysRevLett.86.1078,PhysRevLett.90.076806} and carbon nanotubes~\cite{PhysRevLett.102.036804}, and for the edge-channel spectroscopy in the integer quantum Hall regime~\cite{Altimiras2009,PhysRevLett.120.197701}. In these experiments, EDs were obtained by measuring average current through the tunnel junction (TJ) and through the QD, respectively. As initially understood by Gramespacher and B$\rm\ddot{u}$ttiker~\cite{PhysRevB.60.2375}, energy-resolved local information is accessible even without using any spectral-sensitive detector, being concealed in current fluctuations measured with a local probe rather than in the average current. In this case, energy sensitivity is provided solely by the Pauli exclusion principle, which couples EDs in the equilibrium reservoir and in the studied device in the expression for the shot noise. Noteworthy, there is no external limiting energy scale in this approach besides bath temperature. 

In this paper we demonstrate, to our best knowledge, the first experimental proof of principle of such local noise spectroscopy. To illuminate the main idea of the experiment we first consider two electron reservoirs with EDs~$f_1$ and~$f_2$ coupled, for the moment, by a TJ with an energy-independent transmission probability. The average partial tunneling current in the energy strip~$\delta\varepsilon$ from the $i$-th to the $j$-th reservoir for a single conduction channel is $\delta I_{i\to j}\propto f_i(1-f_j)\,\delta\varepsilon$.
The factors $f_i$ and $(1-f_j)$ are imposed by the Pauli exclusion principle. Still, the expression for the average partial tunneling current through the TJ $\delta I=\delta I_{1\to2}-\delta I_{2\to1}\propto(f_1-f_2)\,\delta\varepsilon$ is identical to the classical case. However, the quasiparticle statistics is revealed in the fluctuations of currents $\delta I_{1\to2}$ and $\delta I_{2\to1}$ which add up in a Schottky-like manner to give the current noise spectral density: $\delta \mysub{S}{I}= 2e\big[|\delta I_{1\to2}|+|\delta I_{2\to1}|\big]\propto\big[f_1+f_2-2f_1f_2\big]\,\delta\varepsilon$. This coupling of the EDs on the two sides of the TJ enables the energy resolution in the shot noise measurement. Provided one takes into account the transmission eigenvalue distribution by the introduction of the Fano-factor in the expression for the shot noise, the above reasoning applies for any multimode conductor with transport occuring at constant energy. Equally important, in the nonequilibrium configuration zero average current through the conductor can coexist with its noise which by far exceeds the Johnson-Nyquist value. For the simplest case of a temperature difference across the conductor this was recently demonstrated for InAs-nanowires~\cite{Tikhonov2016} and for atomic scale junctions~\cite{Lumbroso2018,PhysRevB.98.235432}. 

\begin{figure}[h]
\begin{center}
\vspace{0mm}
\includegraphics[width=0.95\columnwidth]{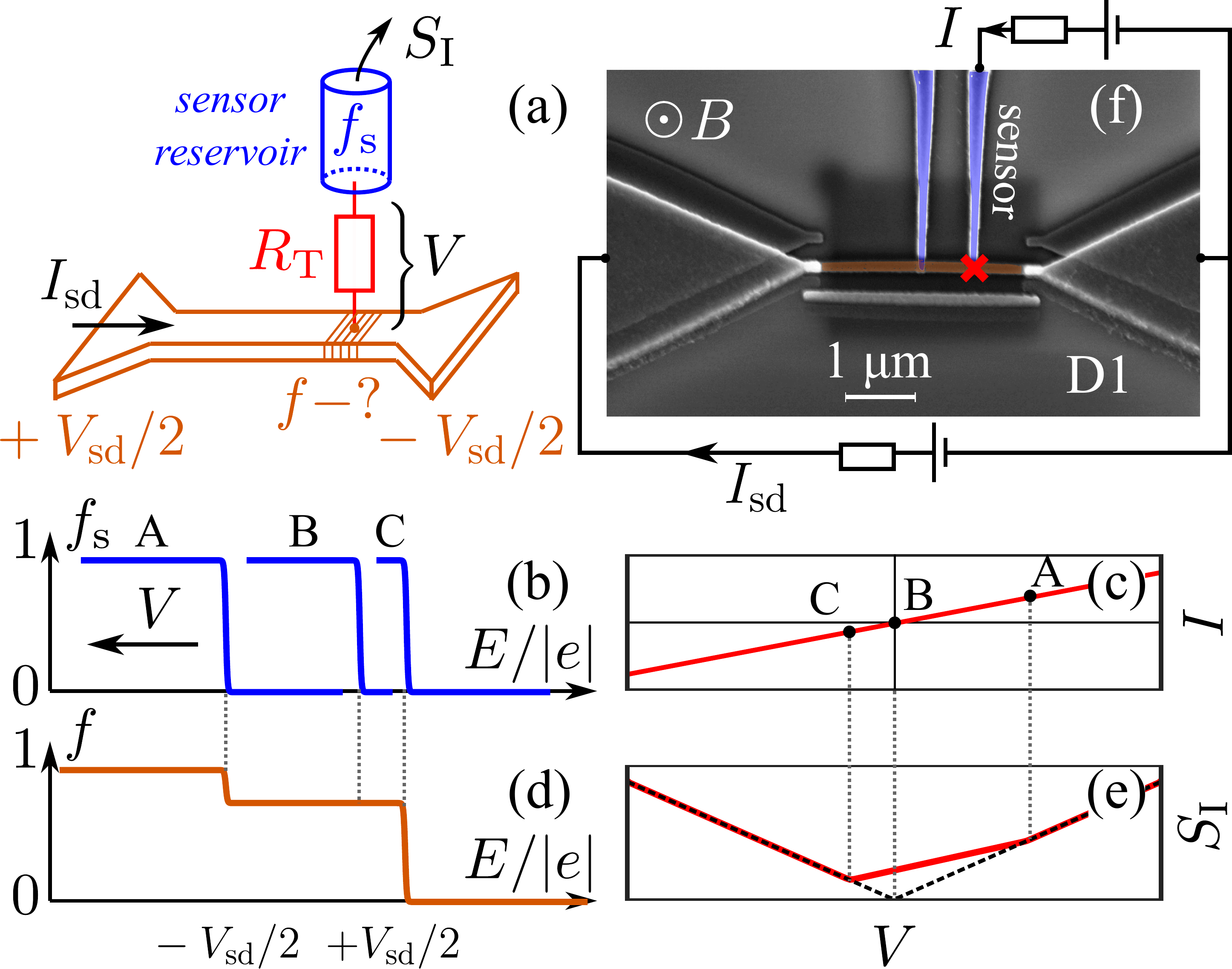}
\end{center}
\caption{(a)~Schematic measurement setup. The central narrow wire is driven out of equilibrium with a transport current~$\mysub{I}{sd}$. Current fluctuations~$\mysub{S}{I}(V)$ are measured from the sensor reservoir which is tunnel coupled to the wire. (b,d)~ED in the sensor~$f_s$ and in the nonequilibrium wire in contact point~$f$ (in case of negligible inelastic scattering at~$T=0$) for~$\mysub{I}{sd}>0$. Bias voltage~$V$ on the TJ effectively scans~$f_{\mathrm{s}}$ relatively to the studied~$f$. (c,e)~Corresponding $I$-$V$ curve and current noise of the~TJ. The thin dotted line shows the standard shot noise curve with equilibrium FD distribution in the wire. (f)~Colored SEM micrograph of the device~D1. The red cross indicates the position of the~TJ used for sensing of ED located at one-quarter distance between two reservoirs.}\label{fig1}
\end{figure}

In the following we will demonstrate the measurements of ED in micrometer-scale nonequilibrium metallic wires, see fig.~\ref{fig1}(a) for the sketch of the experiment setup. The sensor reservoir is described by the equilibrium ED $f_1\equiv f_s=f_0(\varepsilon-eV,T_0)$, and the probed conductor locally by some nonequilibrium~ED~$f_2\equiv f(\varepsilon)$. Here, $f_0(\varepsilon,T)=\big[\exp{\left[\varepsilon/\left(\mysub{k}{B}T\right)\right]}+1\big]^{-1}$ is the Fermi-Dirac (FD) distribution, $V$~is the bias voltage across the TJ connecting the two reservoirs, and~$T_0$ is the bath temperature. Changing~$V$ effectively scans~$f_s$ relatively to the studied nonequilibrium~$f$, see figs.~\ref{fig1}(b) and ~\ref{fig1}(d). The current across the TJ is $I\propto \int\big[f_0(\varepsilon-eV)-f(\varepsilon)\big]\,\delta \varepsilon\propto V$,
see fig.~\ref{fig1}(c). At the same time, the $V$-dependent contribution to the partial current noise spectral density $\delta \mysub{S}{I}\propto f_0(\varepsilon-eV)\big[1-2f(\varepsilon)\big]\,\delta \varepsilon$ leads to the peculiar $\mysub{S}{I}(V)$-dependence with
\begin{equation}
\frac{d\mysub{S}{I}}{dV}\propto 1-2f(eV).
\label{refhere}
\end{equation}
Expression~(\ref{refhere}) is the essence of the local noise spectroscopy we develop here. Analogous expressions, contained already in the pioneering theory of the local noise probe~\cite{PhysRevB.60.2375}, were later derived for the auto-correlation noise~\cite{Kuehne2016} and cross-correlation noise~\cite{Ota2017} in nonequilibrium Tomonaga-Luttinger liquids, for the studies of an ac-biased TJ~\cite{PhysRevB.87.075403}, used for the quantum tomography purposes~\cite{Bisognin2019} and for the semiconducting nanowire based local noise sensors~\cite{Piatrusha2017,Denisov2020}. 
For the nonequilibrium~$f(\varepsilon)$ with possibly many step-like features, each step is  associated with the change of the slope in the~$\mysub{S}{I}(V)$-dependence. Below we explain the relation between these kinks and the local ED in the diffusive wire in present experiment.

Consider, for simplicity, the case of negligible inelastic scattering in the studied wire achieved at low enough~$T_0$. Upon the application of a transport current $\mysub{I}{sd}$ through the wire ED~$f(\varepsilon)$ acquires an intermediate step, see fig.~\ref{fig1}(d) with the height depending linearly on the position along the wire~\cite{Nagaev1992}. At bias voltages~$V$ when the step in FD distribution~$f_s$ crosses the steps in~$f$, the derivative~$d\mysub{S}{I}/dV$ changes its value which is expressed as two kinks in the $\mysub{S}{I}(V)$-dependence, see fig.~\ref{fig1}(e). The presence of electron-electron (\textit{e-e}) scattering in the wire leads to the smoothing of the kinks. 

The colored SEM image of a typical device is presented in fig.~\ref{fig1}(f). The \myq{3}{\mu m}-long \myq{25}{nm}-thick and \myq{100}{nm}-wide copper wire (brown) is evaporated above \myq{20}{nm}-thick Al electrodes (blue) which were first controllably oxidized for \myq{2}{min} with pure oxygen pressure of \myq{1}{mbar} to form a TJ (red cross). The wire is well coupled to two thicker side aluminum reservoirs which are \myq{125}{nm}-thick and were intentionally made as wide as possible~\cite{PhysRevB.59.2871}. These reservoirs are used to turn the wire out of equilibrium with a transport current~$\mysub{I}{sd}$. For sensing purposes, we used Al electrode located either at one-quarter distance between two reservoirs (device~D1) or in the middle of the wire (device~D2). The unused Al electrode on both devices was left unbonded. The typical TJ's resistance is around $25$--\myq{30}{k\Omega} by far exceeding the sum of the wire's resistance (\myq{27}{\Omega}) and its reservoirs resistance ($\approx$\myq{1}{\Omega} to the ground) ensuring negligible heat leakage to the sensor reservoir.
In order to avoid any superconducting effects of Al electrodes, during the noise measurements we applied perpendicular magnetic field at least sufficient to completely suppress superconductivity (\myq{120}{mT}). In the similar fashion we also studied ED in the middle of \myq{3}{\mu m}-long \myq{25}{nm}-thick and \myq{150}{nm}-wide Al wire realized in an all-aluminum TJ device (device~D3, see Appendix~A Fig.~5) with TJ's resistance of \myq{5}{k\Omega} and the wire's resistance of~\myq{10}{\Omega}. 
The noise measurement details, conventional transport and noise properties of the TJs are described in Appendices~A and~B.

\begin{figure}[h]
\begin{center}
\vspace{0mm}
\includegraphics[width=0.9\columnwidth]{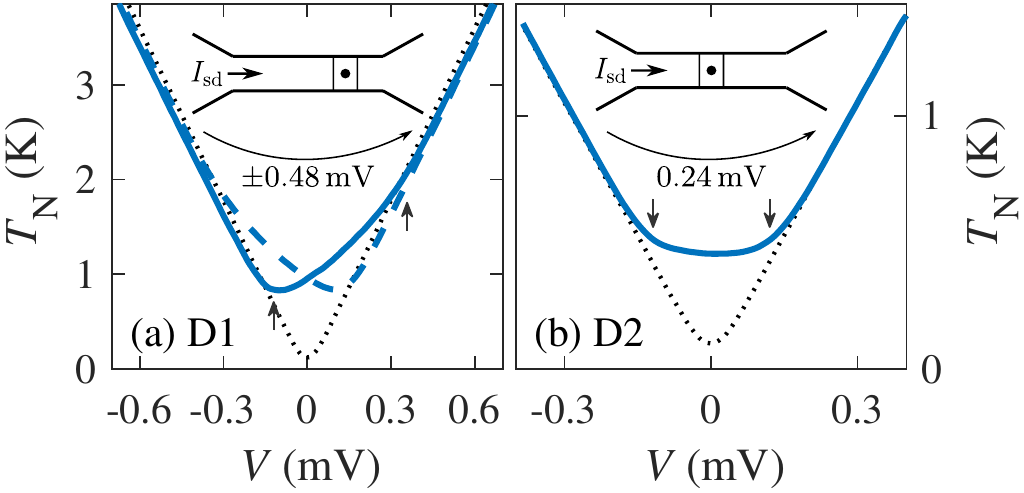}
\end{center}
\caption{Noise temperature vs. bias voltage for the TJs realized at two different positions along the wire at \myvareq{T}{bath}{30}{mK} in a magnetic field of \myvareq{B}{}{5}{T}. (a)~In the asymmetric case, $\mysub{T}{N}$~depends on the polarity of~$\mysub{I}{sd}$ in accordance with fig.~\ref{fig1}(f), reflecting the locality of the extracted~ED. (b)~$\mysub{T}{N}$ is symmetric for the central positioning of the TJ. Dotted lines in both panels demonstrate~$\mysub{T}{N}$ measured at \myvareq{I}{sd}{0}{}.}\label{fig2}
\end{figure}

Knowledge of the Fano-factor of the TJ allows one to infer $f(eV)=1/2-(1/F)d(\mysub{k}{B}\mysub{T}{N})/d(eV)$,
where $\mysub{T}{N}=\mysub{S}{I}\mysub{R}{T}/4\mysub{k}{B}$ is the TJ's noise temperature and $\mysub{R}{T}$ is its resistance. This relation is valid when $\mysub{k}{B}T_0$ is much less than the characteristic energy scale on which the local ED~$f(\varepsilon)$ changes significantly. Fig.~\ref{fig2} demonstrates the dependence $\mysub{T}{N}(V)$ in the devices~D1 and~D2 in large magnetic field~\myq{5}{T}. The dotted lines on the panels~(a) and (b) are measured in the absence of nonequilibrium in the metallic wire, $\mysub{I}{sd}=0$. In this case, the $\mysub{T}{N}(V)$-dependence is typical: it is symmetric with respect to the~$V$ inversion and displays the parabolic transition from the Johnson-Nyquist noise at low~$|V|$ to the linear shot noise at higher~$|V|$.
The finite transport current~$\mysub{I}{sd}$ changes $\mysub{T}{N}(V)$ drastically, see solid lines. For the TJ realized at one-quarter of the way between two reservoirs, fig.~\ref{fig2}(a), $\mysub{T}{N}(V)$ becomes asymmetric with the kink-like features at $V=\mp \mysub{V}{sd}/4$ and $V=\pm 3\mysub{V}{sd}/4$ with upper signs corresponding to $\mysub{I}{sd}>0$ and lower signs corresponding to $\mysub{I}{sd}<0$. These features coincide with the expected kinks positions, as indicated by arrows in fig.~\ref{fig2}(a) for $\mysub{I}{sd}>0$. In the case when the TJ is realized in the middle of the nonequilibrium conductor, fig.~\ref{fig2}(b), the $\mysub{T}{N}(V)$-dependence is symmetric, however with a near-zero bias plateau-like region. Note, how this observation illustrates (\ref{refhere}): scanning the $1/2$-plateau in~$f(\varepsilon)$ with bias voltage~$V$ doesn't change $\mysub{T}{N}$. Here, again, the plateau's boundaries coincide with expected kinks position, see arrows in fig.~\ref{fig2}(b). Overall, Fig.~\ref{fig2} reveals energy and spatial sensitivity of our approach. Note, that similarly looking results were obtained for the noise measurements in a two-terminal TJ under biharmonic illumination~\cite{PhysRevB.87.075403} revealing the effect of interference of the two Fermi seas on both sides of the TJ. The proposed interpretation in terms of EDs, however, is unclear since both Fermi seas are equally important for the creation of the nonequilibrium and are equivalent with respect to the applied ac excitation. On the contrary, in our case nonequilibrium is created exclusively on one side of the TJ while the other reservoir remains in equilibrium and plays the role of a non-invasive sensor.

\begin{figure}[t]
\begin{center}
\vspace{0mm}
\includegraphics[width=\columnwidth]{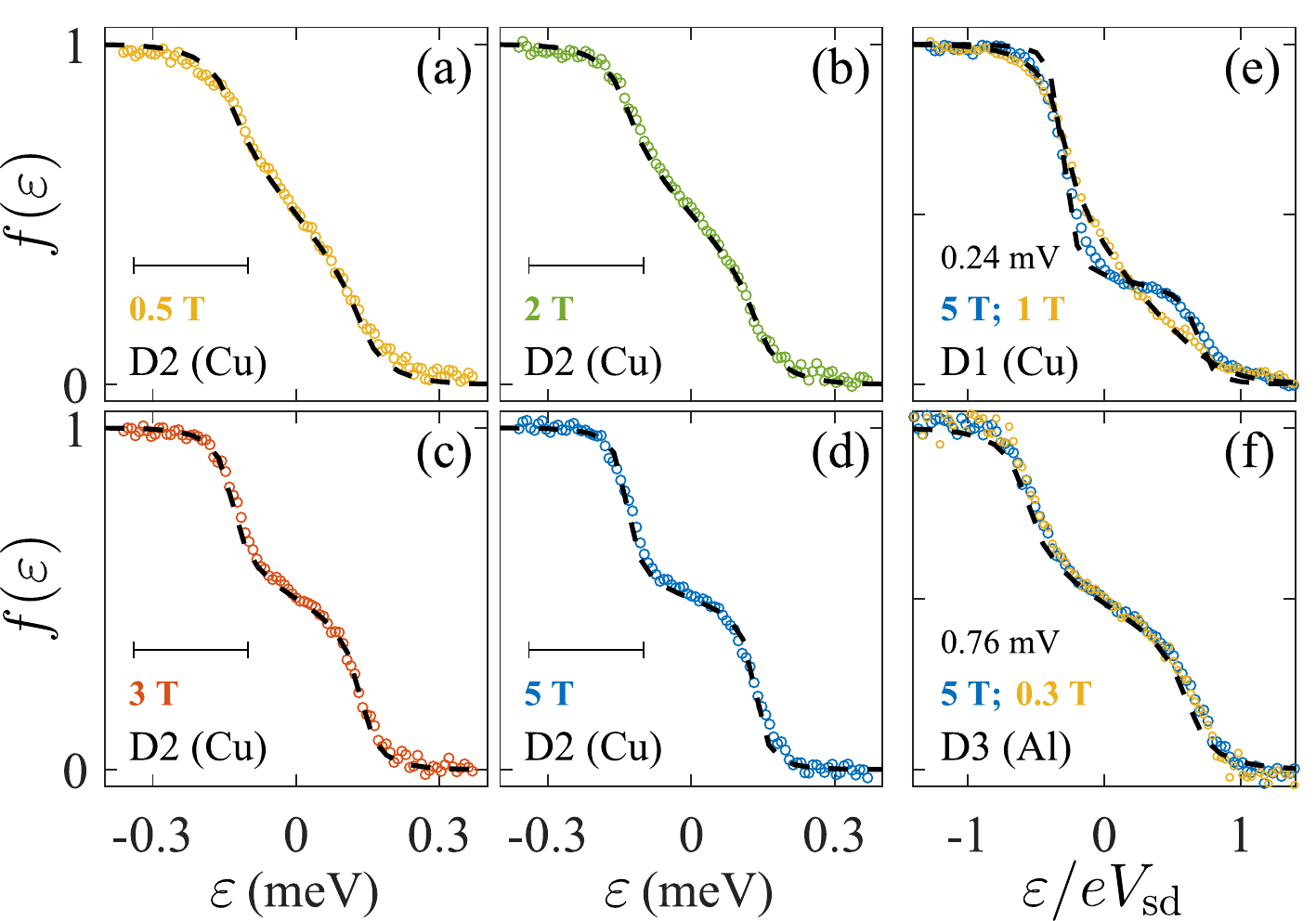}
\end{center}
\caption{(a-d)~Magnetic field evolution of ED in the middle of the Cu wire (device~D2) at~$\mysub{V}{sd}=0.24\,\mathrm{mV}$ (scale bar). Panel~(d) corresponds to the data of fig.~\ref{fig2}(b). The dashed lines are solutions of the Boltzmann equation which fit the experimental EDs best. (e)~ED in the Cu wire (device~D1) at one-quarter distance between two reservoirs at~$\mysub{V}{sd}=0.24\,\mathrm{mV}$ (scale bar). (f)~ED in Al device~D3 at~\myvareq{V}{sd}{0.76}{mV}. All the data are obtained at~\myvareq{T}{bath}{30}{mK}.}\label{fig3}
\end{figure}

For the data similar to that of Fig.~\ref{fig2}, we are able to directly extract the local~ED in the nonequilibrium situation. The results are summarized in Fig~\ref{fig3}. The panels~(a) to (d) present the evolution of ED as a function of the external magnetic field $B$ at \myvareq{I}{sd}{9}{\mu A} with the expected step width \myvareq{V}{sd}{0.24}{mV} drawn by a scale bar. Remarkably, the expected double-step feature in the local ED is clearly seen for $B\geq3$\,T and completely smeared in low $B$-field. This manifests a $B$-dependent thermalization of electrons owing to an inelastic scattering process that can be suppressed by the magnetic field. Such a behavior is not expected for the electron-phonon scattering, which is anyway negligible in our devices up to $\mysub{V}{sd}\approx1\,\text{mV}$ at~\myvareq{T}{0}{30}{mK}, see Appendix~C Fig.~\ref{phonons}. Our result therefore demonstrates the impact of a magnetic field on the \textit{e-e} scattering, which gradually diminishes at increasing $B$. This evolution persists up to~$B\sim5\,\text{T}$, where the effect of magnetic field saturates. Similar behavior is known from~\cite{PhysRevLett.90.076806,PhysRevLett.95.036802}, where it was deduced from the features in differential conductance of a TJ due to Coulomb blockade utilizing the special design of the sensor electrode. The observed behavior is consistent with the presence of dilute magnetic impurities~\cite{PhysRevLett.86.2400,PhysRevB.66.195328} -- impurity-induced energy exchange in small magnetic fields freezes out at increasing~$B$. Alternatively, the similar effect might also result from the presence of paramagnetic oxygen at the copper film surface~\cite{PhysRevB.37.8502}.

Fig.~\ref{fig3}(e) demonstrates EDs at one-quarter distance between two reservoirs in copper device~D1 measured at~\myvareq{T}{0}{30}{mK} for two representative values of a $B$-field. Again, the step-feature is almost indistinguishable in a smaller field \myvareq{B}{}{1}{T}, however the ED is far from the thermal one. As in~D2, ED evolves with increasing~$B$ reaching saturation in~$\sim5\,\text{T}$. In fig.~\ref{fig3}(f) we demonstrate the ED in the middle of aluminum device~D3, similarly obtained at two values of magnetic field. Unlike the case of copper, here the ED is independent of the magnetic field~\cite{anthore:tel-00003518} and is of a double-step form already in small~$B$. We note that it is also observable at higher~\myvareq{T}{bath}{0.5}{K} (see Appendix~D Fig.~\ref{denisov}). While the surface aluminum atoms may form a bath of magnetic moments~\cite{PhysRevLett.112.017001} similar to the copper case, the observed difference between two materials may indicate a smaller density of the magnetic moments and/or their weaker coupling to the conduction electrons in aluminum.

Extracted EDs provide access to \textit{e-e} scattering time in the copper wires as a function of~$B$. The ED inside a quasi-one-dimensional conductor obeys the Boltzmann equation~\cite{PhysRevB.52.4740,PhysRevB.52.7853} $D\partial^2 f(x,E)/\partial x^2+\mysub{I}{coll}(x,E,\{f\})=0$. Here, $D=L^2/\mysub{\tau}{D}$ is the diffusion coefficient, $L$ -- length of the wire, $\mysub{\tau}{D}$ is the diffusion time of electrons along the wire and $x$ is the coordinate along the wire. Taking into account only \textit{e-e}~scattering and assuming the interaction is local, one gets $\mysub{I}{coll}(x,E,\{f\})=\int d\varepsilon\,dE'\,K(\varepsilon)\,f^x_{E'}\left(1-f^x_{E'+\varepsilon}\right)
\times \big[\left(1-f^x_E\right)f^x_{E+\varepsilon}-f^x_E\left(1-f^x_{E-\varepsilon}\right)\big]$, where $K(\varepsilon)$ is the interaction kernel. The dominant role of exchange interaction of electrons with magnetic impurities suggests $K(\varepsilon)=\tau_{ee}^{-1}/\varepsilon^2$, where $\tau_{ee}^{-1}$ is the rate of \textit{e-e} scattering~\cite{PhysRevLett.79.3490,PhysRevLett.86.2400}. Using the numerical relaxation method, we solve the Boltzmann equation and obtain the ratio $\tau_{ee}/\tau_D$ which fits the experimental EDs best. For the copper device~D2 the corresponding best fits are shown by dashed lines in panels~(a-d) of fig.~\ref{fig3}. 

\begin{figure}[h]
\begin{center}
\vspace{0mm}
\includegraphics[width=1\columnwidth]{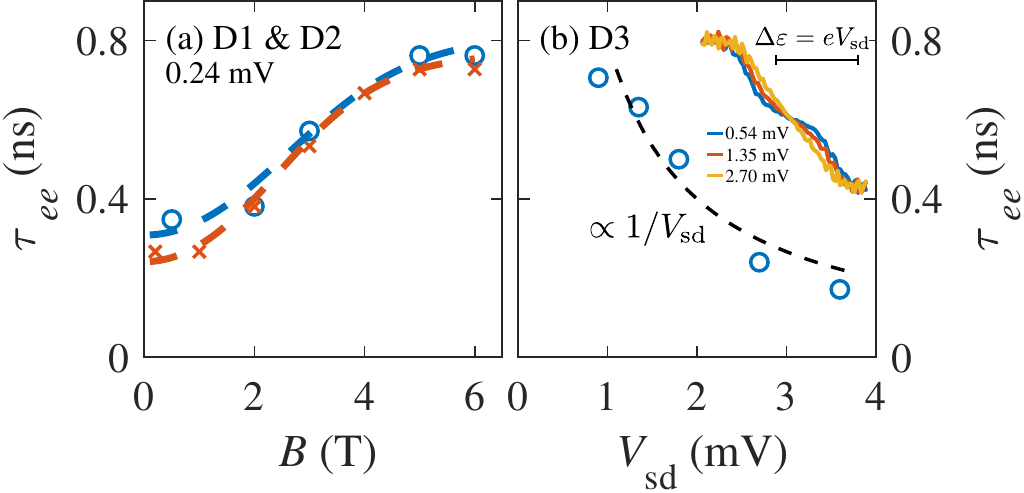}
\end{center}
\caption{(a)~Scattering time as a function of magnetic field in devices~D1 (crosses) and~D2 (circles) at \myvareq{V}{sd}{0.24}{mV}. The dashed lines effective temperatures are $T^*/(e\mysub{V}{sd}/\mysub{k}{B})=0.46$ and $0.39$, correspondingly. (b)~Scattering time as a function of~$\mysub{V}{sd}$ in device~D3 in~\myvareq{B}{}{5}{T}. The inset demonstrates thermalization of ED at increasing bias voltage.}\label{fig4}
\end{figure}

In fig.~\ref{fig4}(a) we plot the obtained $\tau_{ee}$ in dependence of the magnetic field for both copper devices, see the symbols. At increasing $B$ from \myq{0.3}{T} to \myq{6}{T} $\tau_{ee}$ grows monotonically and saturates at high~$B$. This evolution may be understood as follows. For the $B$-dependent \textit{e-e}~scattering rate we assume $1/\tau_{ee}(B)=1/\mysub{\tau}{sf}(B)+1/\tau_0$,
where $\mysub{\tau}{sf}$ is the spin-flip rate due to $B$-dependent scattering involving magnetic impurities, and $\tau_0$ is the $B$-independent scattering rate, e.g., due to the direct Coulomb interaction. For the spin-flip rate, we use the expression similar to that in thermal equilibrium~\cite{PhysRevB.68.085413} $\tau_{\text{sf}}(B)/\tau_{\text{sf}}(B=0)=\sinh\left(g\mu_B B/k_BT^*\right)/(g\mu_B B/k_BT^*)$,
where $\mu_{B}$ is the Bohr magneton, $g$ is the gyromagnetic factor of the magnetic impurities. The effective temperature in our strongly nonequilibrium case should scale with the bias voltage on the metallic wire $T^*\sim e\mysub{V}{sd}/\mysub{k}{B}$. This expression closely describes our data, assuming $g=2$, $T^*/(e\mysub{V}{sd}/\mysub{k}{B})=0.46$ and $0.39$ and $\tau_0/\mysub{\tau}{D}=1.02$ and $0.96$, respectively, for the devices D1 and D2. We compare the obtained values for $\tau_{ee}$ with the theoretical prediction of~\cite{Altshuler_1982,PhysRevB.65.115317} in Appendix~E.

We note that, experimentally, the double-step feature smooths out at increasing $\mysub{V}{sd}$. This is illustrated on the inset of fig.~\ref{fig4}(b), where EDs measured in~D3 are plotted as functions of the normalized energy $\varepsilon/(e\mysub{V}{sd})$ for various values of $\mysub{V}{sd}$. Smoothing of EDs is an obvious consequence of the direct Coulomb interaction which starts to dominate at increasing excess quasiparticle energy. For the kernel of Coulomb interaction $\mysub{K}{Coulomb}(\varepsilon)\propto\varepsilon^{-3/2}$ $\tau_0$ depends on the exact ED~\cite{PhysRevB.65.115317} which, in turn, depends both on the energy of the quasiparticle and on the position along the wire. To estimate $\tau_{ee}$ we formally use the same kernel as before, however, with the bias-dependent scattering time $K(\varepsilon)=\tau_{ee}^{-1}(\mysub{V}{sd})/\varepsilon^2$. For device~D3, the results are shown in fig.~\ref{fig4}(b). The dependence $\tau_{ee}(\mysub{V}{sd})$ is stronger than the expected in 1D~\cite{PhysRevB.65.115317} $\tau_{ee}\propto \mysub{V}{sd}^{-1/2}$, probably indicating the transition of the wire to effectively larger dimensionality in terms of energy relaxation.

We note that the demonstrated approach is also applicable to the study of nonequilibrium configurations associated with spin (or valley, etc.) currents. Naturally, in this case the local noise probe should additionally conserve the respective quantum number. Theoretically, it is known that the current noise reflects the degree of spin imbalance in the reservoirs~\cite{PhysRevB.84.073302}. Experimentally, this concept was recently investigated in the study of the spin accumulation driven shot noise across a tunnel barrier with a spin polarized injection contact~\cite{PhysRevLett.114.016601}. In this respect, potentially, our approach may be useful for investigating the microscopic details of spin (or valley, etc.) relaxation. 

In summary, we experimentally demonstrated the local sensing of the nonequilibrium ED in a diffusive metallic wire based on the shot noise measurements with a tunnel junction. This approach relies solely on the Pauli exclusion principle and works in the absence of the energy-selective features in conductance. Consequently, the energy resolution of such a measurement is only limited by the bath temperature. The spatial resolution is virtually unlimited with state-of-the-art noise scanning techniques~\cite{PhysRevLett.75.1610,Kemiktarak2007,Massee2018,PhysRevB.100.104506}. The approach is quite universal and equally suitable for the measurements of the nonequilibrium configurations created by charge, spin~\cite{Khrapai2017} and valley etc. currents, hence paving the way for the realization of existing~\cite{PhysRevB.60.2375,Kuehne2016,Ota2017,eTikhonov2016} and various novel local noise probes.

Development of local noise measurements, measurements in devices~D1 and D2 and analysis of ED evolution in magnetic field were performed under the support of Russian Science Foundation Grant No. 18-72-10135.
Measurements in device~D3 were performed under the support of the Russian Science Foundation Grant No. 19-12-00326. Fabrication of device~D3 was performed using equipment of MIPT Shared Facilities Center and with financial support from the Ministry of Science and Higher Education.
We thank H.\,Pothier, I.L.\,Aleiner and A.D.\,Zaikin for helpful discussions.

\appendix
\section{Device and measurement techniques}
\begin{figure}[h]
	\begin{center}
		\includegraphics[width=.65\columnwidth]{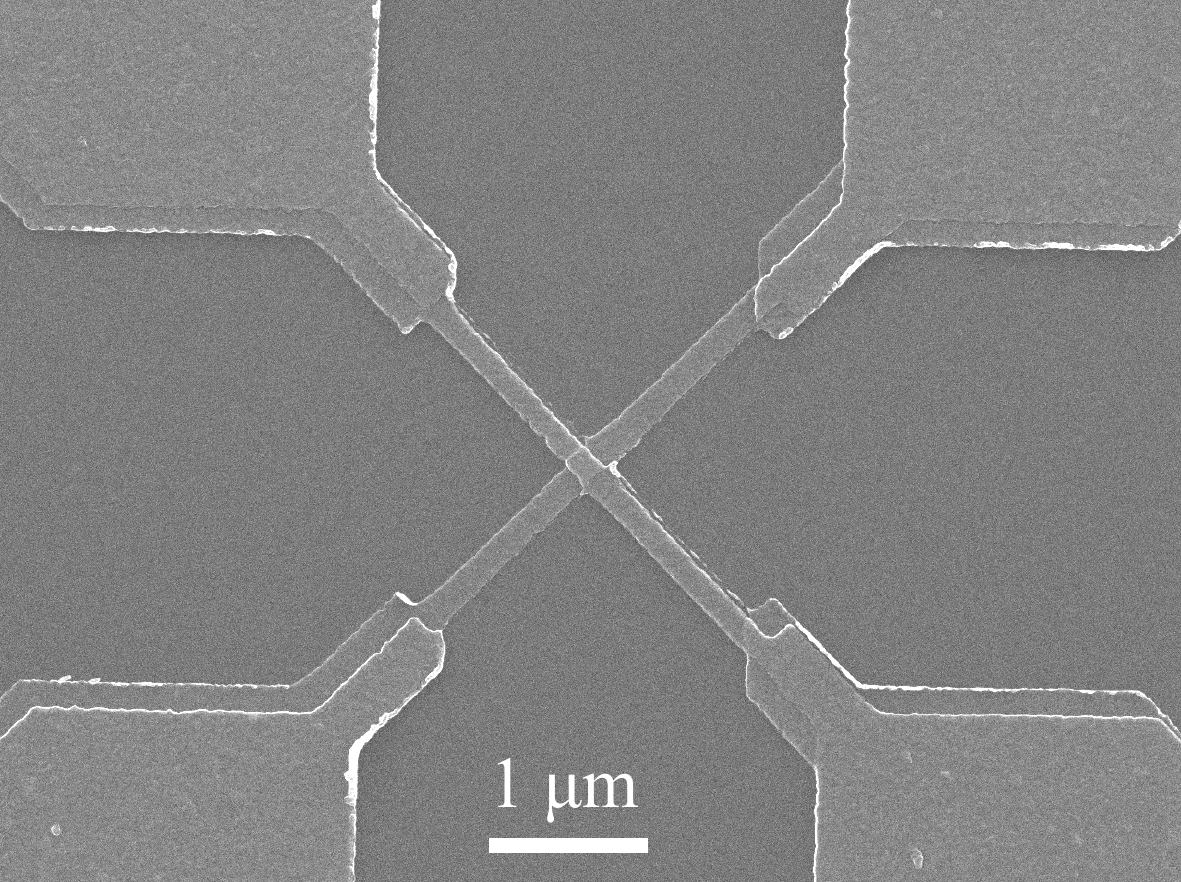}
	\end{center}
	\caption{\textbf{SEM microphotograph of the device D3.} All-aluminum TJ (in the middle) is realized between \myq{3}{\mu m}-long \myq{25}{nm}-thick and \myq{150}{nm}-wide Al wires.}
	\label{khrapach}
\end{figure}

To characterize our devices in terms of the electronic elastic mean free path (mfp) we do as follows. The diffusion coefficient~$D$ is first obtained from the Einstein's relation $\sigma=\nu_{\text{F}}e^2D$. Then, the mfp is obtained from $D=1/3v_{\text{F}}l$ with $v_{\text{F}}$ the Fermi velocity. For Cu devices D1 and D2 we find $D=120\,\mathrm{cm^2/s}$, $\tau_D=0.8\,\text{ns}$, $l_{\text{mfp}}=23\,\text{nm}$;
for Al device D3 -- $D=200\,\mathrm{cm^2/s}$, $\tau_D=1.2\,\text{ns}$, $l_{\text{mfp}}=30\,\text{nm}$.

\vspace{5mm}
The noise spectral density was measured using the home-made low-temperature amplifier (LTamp) with a voltage gain of about \myq{10}{dB} and the input current noise of $\sim 2$--\myq{6\times10^{-27}}{A^2/Hz}. The voltage fluctuations on a \myq{6.4}{k\Omega} load resistor were measured near the central frequency \myq{7}{MHz} of a resonant circuit at the input of the LTAmp. The output of the LTamp was fed into the low noise \myq{75}{dB} gain room temperature amplifcation stage followed by a hand-made analogue filter and a power detector. The setup was calibrated using the equilibrium Johnson-Nyquist noise thermometry. Unless otherwise stated, the measurements were performed in a cryogen free Bluefors dilution refrigerator BF-LD250 at a bath temperature of \myq{30}{mK}.
\section{Transport and noise properties of the TJs}
\begin{figure}[h]
	\begin{center}
		\includegraphics[width=1\columnwidth]{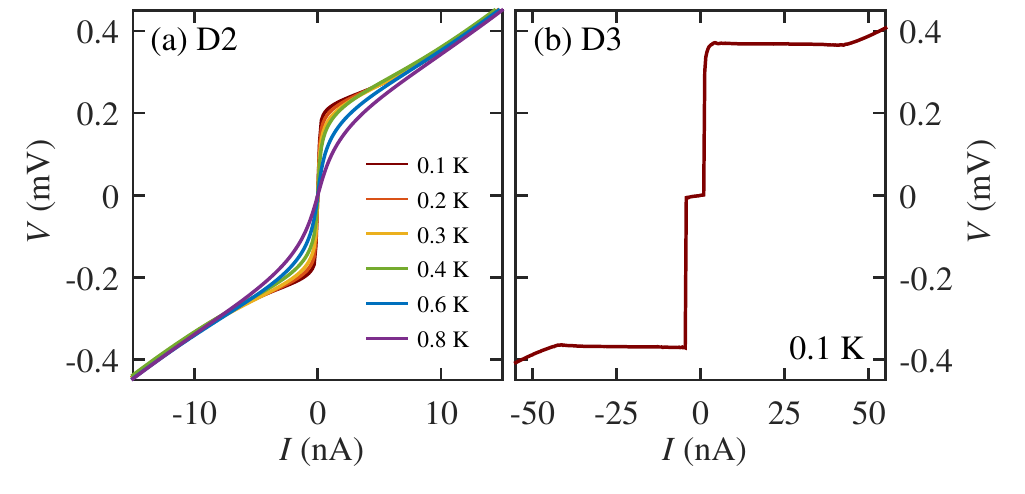}
	\end{center}
	\caption{\textbf{$I$-$V$ curves of the copper and aluminum devices.} (a)~In~D2, the $I$-$V$ characteristics of the TJ in~$B=0$ demonstrates the typical NIS behavior with drastic decrease of subgap conductance with decreasing temperature, and conductance peaks at $\sim$\myq{190}{\mu eV}, reflecting the maxima in the density of states of the superconducting~Al. (b)~In~D3, the $I$-$V$ curve in~$B=0$ demonstrates the typical SIS~behavior.}
	\label{ivs}
\end{figure}

\begin{figure}[h]
	\begin{center}
		\includegraphics[width=1\columnwidth]{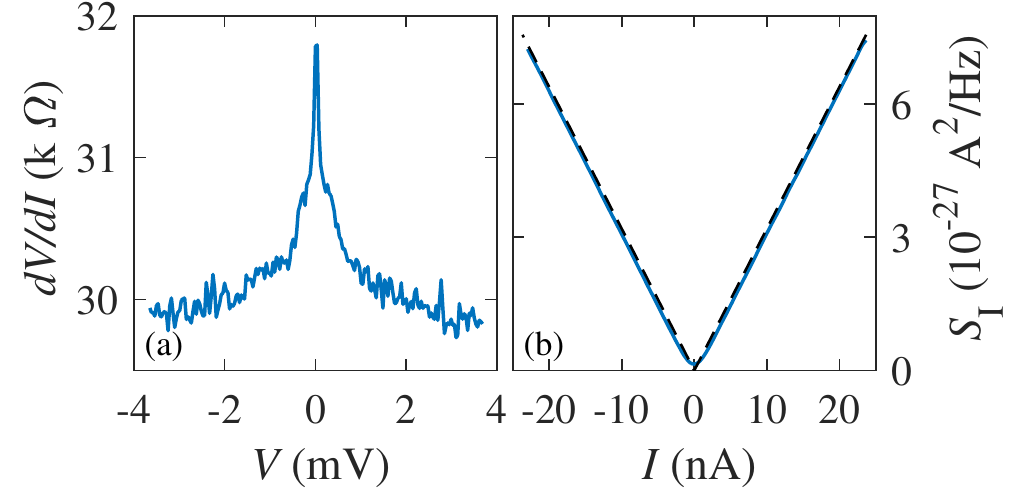}
	\end{center}
	\caption{\textbf{Differential resistance and shot noise of the tunnel junction.} (a)~Differential resistance and (b)~shot noise of the tunnel junction in the device~D1 measured at \myvareq{T}{bath}{30}{mK} in \myvareq{B}{}{0.3}{T}. The nonlinearity of $dV/dI$ is approximately $5\%$.}
	\label{zai}
\end{figure}

For all three devices we first characterize the TJs in terms of conventional transport and noise properties. For devices D1 and D2 the $I$-$V$ characteristics in $B=0$  demonstrated the typical normal metal-insulator-superconductor (NIS) behavior with drastic decrease of subgap conductance with decreasing temperature, and conductance peaks at $\sim$\myq{190}{\mu eV}, reflecting the maxima in the density of states of the superconducting~Al. The device~D3 showed the typical SIS~behavior (see Fig.~\ref{ivs} for the $I$-$V$ curves of both devices). All three devices demonstrated almost linear $I$-$V$~curves in finite magnetic field suppressing superconductivity with negligible conribution of interaction effects~\cite{Zaikin2019} (see Fig.~\ref{zai}). In terms of noise in the normal state, devices~D1 and D3 demonstrated the standard Fano-factor~$F=1$, common for TJs. In device D2 we measured a linear behavior typical for TJs, however, with $F=0.6$ which might be a result of a pinhole. This junction also acts as a current noise-to-ED converter, yet with a slightly smaller sensitivity owing to the reduced shot noise.

\section{Phonons}
\begin{figure}[h]
	\begin{center}
		\includegraphics[width=.6\columnwidth]{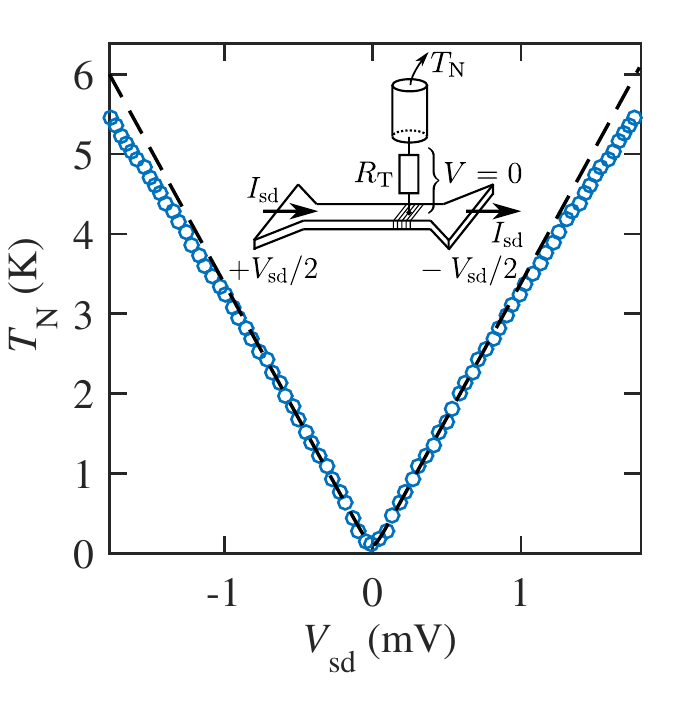}
	\end{center}
	\caption{\textbf{Local noise measurement in the Cu strip.} The inset shows the measurement scheme. Data in $0.3\,\text{T}$ and in $6\,\text{T}$ are almost indistinguishable.}
	\label{phonons}
\end{figure}

The linear dependence~$T_{\text{N}}(V_{\text{sd}})$ at $V_{\text{sd}}\lesssim1\,\text{meV}$ at $T_{\text{bath}}=30\,\text{mK}$ (see Fig.~\ref{phonons}) demonstrates the absence of \textit{e-ph}~energy relaxation at corresponding excess energies of quasiparticles (qp). At higher qp energies the~$T_{\text{N}}(V_{\text{sd}})$-dependence becomes sublinear indicating the power flow from electron system to the phonon one.  
\section{Average and local noise measurements in the Al strip at $0.56\,\text{K}$}
\begin{figure}[h]
	\begin{center}
		\includegraphics[width=1\columnwidth]{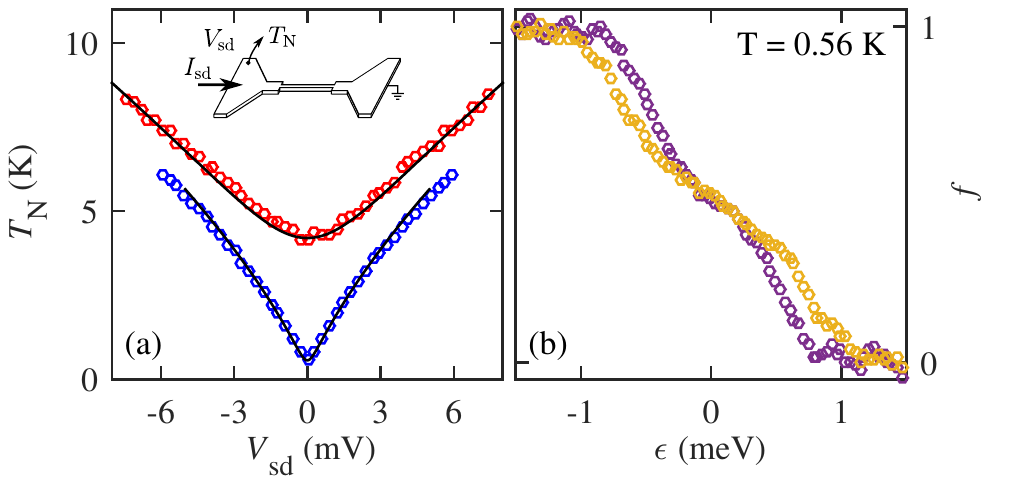}
	\end{center}
	\caption{\textbf{Average and local noise measurements in the Al strip.} (a)~(Symbols) Average noise temperature as a function of bias voltage across the heater at~$T=4.2~\text{K}$ (red) and $T=0.56~\text{K}$ (blue). Numerical simulation taking into account geometry of the sample is shown by solid lines. (b)~ED in Al wire at $T=0.56\,\text{K}$ at $V_{\text{sd}}=1.2\,\text{mV}$ (violet) and $V_{\text{sd}}=1.8\,\text{mV}$ (yellow).}
	\label{denisov}
\end{figure}

Numerical simulation taking into account geometry of the sample fits experimental data, see Fig.~\ref{denisov}(a), for $\Sigma_{\text{e-ph}}=2.3\times10^{11}~\mathrm{W/m^3K^3}$. This value allows one to estimate the \textit{e-ph} scattering length 
$l_{\text{e-ph}}=\sqrt{(\sigma {\cal L})/(3\Sigma_{\text{e-ph}}T)}$
to be $2.3\,\mu\text{m}$ at $T_{\text{bath}}=0.56\,\text{K}$ which is only slightly less than the length of the constriction $l=3\,\mu\text{m}$. This fact allows the observation of double-step feature at $T_{\text{bath}}=0.56\,\text{K}$ as shown in Fig.~\ref{denisov}(b). In the simulation electronic heat conduction is assumed to satisfy the Wiedemann-Franz law $\varkappa=\sigma {\cal L} T$, where ${\cal L}=(\pi^2/3)(k_B/e)^2=2.44\times10^{-8}\,W\Omega K^{-2}$ is the Lorenz number. 
\section{Energy relaxation time estimation}
According to~\cite{PhysRevB.65.115317}, energy relaxation time in a 1D case is given by
\begin{equation*}
\frac{\hbar}{\tau_E}=\frac{e^2}{\hbar}\frac{L_{\varepsilon}}{\sigma_1}\varepsilon,\quad L_{\varepsilon}=\sqrt{\frac{\hbar D}{\varepsilon}},
\end{equation*}
where $\sigma_1$ is the 1D conductivity. For our copper wires, using $D=120\,\mathrm{cm^2/s}$ and $\sigma_1=10^{-7}\,\text{m}/\Omega$, we estimate (at~$\varepsilon=0.24\,\text{meV}$)
\begin{equation*}
\tau_E\approx 6\,\text{ns},\quad L_{\varepsilon}\approx200\,\mathrm{nm}.
\end{equation*}
The contribution from the triplet channel (spin density fluctuations) is practically of the same value and may further decrease~$\tau_E$, making it comparable to the experimental value.
\section{Applicability of the approach}
Overall, the data presented in the main text evidence the power of the local shot noise measurement for the energy resolution of the electronic states out of equilibrium. There are two necessary conditions for this approach to work. One is the elasticity of charge transport through the TJ, which would then preserve spectral information. The second one requires the much smaller thermal conductance of the TJ compared to that of the studied conductor~\cite{Tikhonov2016,Piatrusha2017,Denisov2020,eTikhonov2016,larocque}, similarly to the analogous electrical requirement for conventional voltmeters. To probe low-resistance conductors, these conditions, alongside with TJs, are fulfilled for elastic InAs nanowire-based sensors, allowing additionally thermoelectric or spin-to-charge conversion studies~\cite{eTikhonov2016,Khrapai2017}.

%
\end{document}